\documentclass[preprint,pre]{revtex4}

\usepackage{graphicx}

\usepackage{enumerate}
\usepackage[monochrome]{color}

\usepackage{amssymb,amsfonts,amsmath,color}

\begin{document}

\title{Environmental stochasticity and the speed of evolution}

\author{Matan Danino}
\affiliation{Department of Physics, Bar-Ilan University,
Ramat-Gan IL52900, Israel}

\author{David A. Kessler}
\affiliation{Department of Physics, Bar-Ilan University,
Ramat-Gan IL52900, Israel}

\author{Nadav M. Shnerb}
\affiliation{Department of Physics, Bar-Ilan University,
Ramat-Gan IL52900, Israel}

%%%%%%%%%%%%%%%%%%%%%%%%%%%%%%%%%%%%%%%%%%%%%%%%%%%%%%%%%%%%%%%%

\begin{abstract}
\noindent Biological populations are subject to two types of noise: demographic stochasticity due to fluctuations in the reproductive success of individuals, and environmental variations that affect coherently the relative fitness of entire populations. The rate in which the average fitness of a community increases has been considered so far using models with pure demographic stochasticity; here we present some theoretical considerations and numerical results for the general case where environmental variations are taken into account. When the competition is pairwise, fitness fluctuations are shown to reduce the speed of evolution, while under global competition the speed increases due to environmental stochasticity.
\end{abstract}

\maketitle

\section{Introduction}

Life forms evolve via a continuous process of competition and selection. Malthusian dynamics drives any population to its abundance limit, set by the carrying capacity of its environment and interaction with other populations. In this state the process becomes a zero-sum game and the relative fitnesses of individuals govern their chance to stay alive and to reproduce. When the fitness  is an inherited feature, as in the case where it reflects the characteristics of a genotype, a new strain (species, haplotype) appears each time a significant mutation happens. The abundance of strains with beneficial mutations and higher fitness grows, on average, at the expense of inferior strains. If the supply of beneficial mutations is unlimited, this process  leads to a continuous increase in the average fitness of the whole community.

The rate in which the average fitness increases has been considered by many authors \cite{tsimring1996rna,rouzine2003solitary,park2007clonal,desai2007speed}. It was shown that the averaged fitness grows linearly in time, and the dependence of its speed on the problem parameters, such as the community size $N$, the rate of mutation $\nu$ and the average strength of a single mutation $\Delta s$, was analyzed. However, in all these studies the absolute fitness of an individual (and its non-mutant lineage) is considered independent of time. The aim of this paper is to study the case where this assumption is broken, and in particular to consider the effect of stochastic environmental variations that may lead to temporal fluctuations of the selective force.

Let us envisage a  community of size $N$ composed of  many strains/species of abundances $n_1, \  n_2$ and so on ($\Sigma_i n_i = N$), when these species are playing a zero-sum game. Competition takes place by choosing one individual to die and recruiting an offspring of another individual in its place, where the relative fitnesses of the players govern the probabilities of these two choices. The randomness involved in such a process is known as \emph{demographic stochasticity} and is attributed (like shot noise) to the fact that individuals are discrete. If two species have the same abundance $n$ and the same fitness,  after one generation (i.e., after $N$ elementary birth-death events) their abundances will differ, typically, by  ${\cal O}(\sqrt{n})$.

However, in reality, it is often the case that the relative fitness of a certain genotype  varies through time~\cite{bell2010fluctuating,bergland2014genomic}. Environmental variations and changes in the composition of the community may alter the usefulness and effectiveness of different traits, so the effective fitness of a species  fluctuates. Models in which the fitness landscape changes over time were suggested by some authors in order to explain the ability of evolutionary pathways to cross fitness valleys~\cite{gavrilets2010high}, and the importance of such models for the understanding of the evolutionary process in general was highlighted recently by~\cite{messer2016can}. The response of microorganism communities and their evolutionary dynamics to fitness fluctuations has been considered experimentally by a few groups~\cite{dean2017fluctuating,steinberg2016environmental,taute2014evolutionary}.

In this paper we would like to address the speed of evolution problem in the presence of fluctuating selection. We will consider a case where the time-average of the (logarithmic) fitness of a species is $s_0$, while the instantaneous fitness fluctuates, and is given by $s(t) = s_0 +  \eta(t)$, where the (zero mean) noise $\eta(t)$ reflects environmental variations (or any other natural factor) that change the fitness of a certain genotype~\cite{engen2014evolution, saether2015concept, cvijovic2015fate}.

During evolution, mother and daughter species have many traits in common, and one should make a distinction between environmental variations that affect both of them in the same way and variations that modify the relative fitness of one of them with respect to the other.  Here we study  the latter case and consider only the differential response of the species to environmental stochasticity, with an emphasis on selective forces that reverse their sign through time. In the same spirit, when many species are considered, $\eta(t)$ is taken to be an IID random variable chosen  independently for each species. Such a model may be seen as a first step towards a more realistic theory, in which closely related species have stronger niche overlap and more coherent response to environmental variations.

The environmentally induced fitness fluctuations are characterized by their amplitude $\gamma$ and by the persistence time of the environment, $\delta$, measured in units of generations.  If the persistence time is, say, one generation (i.e., $\delta =1$), the abundance of two species that have the same time-averaged fitness $s_0$ and the same initial abundance $n$ will differ, after one generation, by $\gamma n$, so  environmental stochasticity  is much stronger than demographic noise when the abundances are large \cite{lande2003stochastic}.

Before we consider a specific model, we would like to emphasize a crucial observation of past studies. Naively, one would expect that environmental stochasticity  destabilizes the dynamics of a community, making fluctuations larger and  driving species to extinction more frequently. However, as pointed out by Chesson and coworkers \cite{chesson1981environmental,hatfield1989diffusion,chesson1994multispecies}, under some circumstances environmental stochasticity acts as a stabilizer and supports species coexistence. To account for that, we consider here two generic models: one for environmental stochasticity without such a noise-induced stabilizing mechanism (model A), the other with this mechanism (model B).

In the next section we will describe these two processes in some detail and explain the way we incorporate them into a model of evolutionary dynamics.  In section III we define both models together and sketch the numerical procedures we have used to simulate them. A review of the known results for the pure demographic noise case is given in section IV. In this section we emphasize the distinction between the successional-fixation (selective sweeps)  and the clonal interference phase and that between the regime of weak selection and the regime of strong selection. Section V contains our main results for model A and model B, with analytic expression for the successive-fixation phase and numerical simulations for both that phase and the clonal interference phase. The  fundamental theorem of natural selection \cite{fisher1930genetical,frank1992fisher} is discussed in that context.  Finally we provide a discussion of the main outcomes of our work and point out a few future directions.

\section{Two models of environmental stochasticity}

Environmental variations may affect the fitness of strains or species via many different mechanisms that have to do with many traits. To capture the essence of the problem, we study here  two simple zero-sum games for asexually reproductive individuals, where the fitness determines the chance of reproduction.

We consider a closed system that allows for $N$  individuals. The instantaneous state of this system is fully characterized by a list of strain abundances and their relative fitnesses: the $n_1$ individuals in strain 1 have relative fitness $s_1$ and so on. Only fitness differences ($s_i - s_j$)  play a role in the instantaneous  dynamics.  Relative fitness is measured from the mean value of the fitness in the population $\bar{s} = \sum_i s_i n_i$, and  the speed  of evolution $v_{ev}$ is defined to be $d\bar{s}/dt$. $\bar{s}$ is the mean of the individual fitness probability density, detailing how many individuals have relative fitness between $s$ and $s+ds$.

The shape of the fitness density histogram (for fixed $N$) reflects the balance between selection and mutation. Selection constantly drives low-fitness species to extinction and therefore reduces the variation in fitness, while mutation generates more and more new strains and broadens the fitness distribution.  In the long run the fitness distribution converges to a stable shape of a soliton \cite{rouzine2003solitary} that moves at constant speed $v_{ev}$. In theories without environmental stochasticity $v_{ev}$ depends on the width of the soliton: when the width is large a new mutant at the leading edge of the soliton grows faster (since the fitness difference between its fitness and $\bar{s}$ are larger) and the speed increases. Fisher's  fundamental theorem of natural selection \cite{fisher1930genetical,ewens1989interpretation} states that the instantaneous  speed of evolution is \emph{equal} to the instantaneous genetic variance in fitness.

Now let us describe  two different ways to incorporate fluctuating selection in such a system.

\begin{itemize}
  \item Model A (no stabilizing effect): Every elementary competition event is a duel: two individuals out of $N$ are chosen at random (without replacements) and  fight against each other. The loser dies, the winner produces a single offspring.  The chance of an individual to win a duel depends on its relative fitness with respect to its competitor. If individual 1 belongs to species $i$ and its fitness is $s_i$,  while  2  belong to species $j$ with  $s_j$, 1 wins with probability $P_1$ that depends on $s_i-s_j$, and the chance of 2 to win is $1-P_1$. In model A $P_1$ is given by,
      \begin{equation} \label{maritan}
      P_1 = \frac{1}{1+e^{s_i - s_j}} \approx \frac{1}{2} + \frac{s_i - s_j}{4},
      \end{equation}
      where the last term provides a decent approximation when $|s_i - s_j| \ll 1$.

In model A the competition is pairwise. It may describe, for example, the dynamics of animals on an island where a random encounter between two individuals may end up in a life and death battle over food, a mate or a territory, or the process of local competition for space between two strains of bacteria~\cite{crowley2005general,lloyd2015competition}.

\item Model B (with stabilizing effect): In this model the competition is global (or free for all), and any birth-death event involves all the individuals in the community. First, one individual is chosen at random (independent of its fitness) to die. Then, the chance of species $k$ to win the open ``slot", $P_k$, is given by
    \begin{equation} \label{moran}
     P(k) =  \frac{n_k e^{s_k}}{\sum_j n_j e^{s_j}}.
     \end{equation}

    Model B may describe the dynamics of, say, a forest with many species of trees and long-range seed dispersal. The seed bank in the soil reflects the composition of the whole community  and upon the death of an adult tree one of the local seeds is chosen to capture the newly opened gap with a chance proportional to its fitness. Similarly, it models the competition for a common resource between bacterial strains if the diffusion constant of the resource is large. In model B the overall chance of a species to increase its population reflects both its abundance and its instantaneous fitness. Eq. (\ref{moran}) is used by many authors to describe a Moran process (see, e.g., \cite{haeno2013stochastic,dean2017fluctuating}) and the  infinite-$N$ (deterministic) limit of this dynamics is described by  the replicator equation.
\end{itemize}

In the ``neutral" case, i.e., when there are no fitness differences between individuals, model A and model B coincide since they both yield the same transition probabilities. However, when fitness varies in time the two models differ strongly \cite{danino2016stability,danino2016effect,hidalgo2017species}. For a two-species competition under model A, for example,  the time to absorption (the time until one of the species goes extinct and the other captures the whole system) scales, as $N \to \infty$,  like $\ln N/|\Delta s|$, where $\Delta s$ is the mean fitness difference between the two species.  On the other hand under the dynamics of model B the time to absorption scales like  $N^{(1- |\Delta \tilde {s}|)/\delta}$, where $\Delta \tilde {s} \equiv 2 \Delta s/\gamma^2$ is the effective strength of selection in this system \cite{danino2016stability}. As long as $|\Delta \tilde {s}| <1$, model B is more stable than model A due to noise-induced stabilization.

To get some intuition for the mechanism that allows the stochasticity to act as a stabilizing factor in model B, let us think about a  ``winner takes all" version of this model with two species and ten individuals. At each year one individual is picked at random to die, and one species is chosen at random to win the empty slot. Starting with 8 red individuals and 2 green, when the environment favors the greens their chance to increase their abundance by one is 0.8, but when the reds are favorable  their chance to grow  is only 0.2.  \emph{Rare species have a larger chance to grow in abundance just because they are rare}, and this implies that the noise acts to stabilize the 50:50 state and to facilitate the invasion of rare species. This effect survives even if one of the species has some average fitness advantage, as shown in \cite{chesson1981environmental,hatfield1989diffusion,danino2016stability,danino2016effect}

Model A does not support this stabilizing effect.  If the dynamics of the system described above takes place in a series of duels, starting from 8 reds and 2 greens, the chance for an interspecific duel will be $32/90$,  so the probabilities to end up at 9:1 or at 7:3 will be the same: $16/90$, meaning that there is no preference for rare or common species and no stabilizing effect.

Another fundamental  difference between model A and model B appears when the persistence time of the environment vanishes,  $\delta \to 0$. Model A reduces, in such a case, to a model with pure demographic noise since one picks the relative fitness independently for each elementary duel. On the other hand, in model B the stabilizing effect is maximal when $\delta \to 0$, since, as we have seen, the effect occurs on the level of an elementary event. When $\delta$ is large, on the other hand, the stabilizing effect disappears \cite{chesson1981environmental,danino2016stability}.

Our main goal in what follows is to explain the effect of stochasticity on the evolutionary dynamics with and without a stabilizing mechanism. We begin with an exact definition of the different dynamics through the  numerical procedures that we have used to simulate them, and by stating the main  known results about a two-species system.

\section{Model definitions and numerical procedures}

 We have simulated a community of $N$ individuals that may belong to different species, where the (log) fitness of each species or strain $i$ is characterized by two numbers: its time-averaged value $s_0^i$ and its instantaneous value $s^i(t)$. In the initial state all individuals except one belong to species 1 with average  fitness $s^1_0 = 0$, one individual belongs to species 2 with average fitness $s^2_0 = \Delta s$.

 Environmental variations are responsible to the difference between $s^i_0$ and $s^i(t)$, and are characterized by the strength of variations $\gamma$ and the persistence time of the environment $\delta$. After each elementary birth-death step ($1/N$ generations) the environment flips with probability $1/(N \delta)$, so its persistence time is taken from a geometric distribution with mean $\delta$ generations. After each shift of this type the instantaneous fitness of the $i$'s species is set to be $s^i = s_0^i + \eta^i$, where $s^i_0$ is a species specific, time-independent parameter and   $\eta^i$ is picked, independently for any $i$, from a uniform distribution on the interval $[-\sqrt{3} \gamma, \sqrt{3} \gamma]$, so that $Var(\eta^i) =\gamma^2$.

In each elementary step one individual is chosen at random. With probability $\nu$ this is a mutation step and the mutant becomes the originator of a new species, say $k$.  In such a case both the mean fitness $s^k_0$ and the instantaneous fitness $s^k(t)$  of the chosen individual  increase over its mother species by  $\Delta s$ with probability $1/2$ or decrease by  $\Delta s$ with probability $1/2$. With probability $1-\nu$ we have a  competition step, and for this case the algorithms of model A and model B are different.

\begin{description}
  \item[Model A ] In a competition step, another (different) individual is chosen at random to compete with the already chosen one. The chance of an individual $i$ to win the competition against $j$ is given by Eq. (\ref{maritan}). An offspring of the winner, that inherits its parent fitness (both the  average $s_0$  and the instantaneous fitness $s_0 + \eta(t)$) then replaces the loser.
  \item[Model B] In a competition step, the chosen individual is removed. It is replaced by an offspring that belongs to species $k$ with the probability $P_k$ given in Eq. (\ref{moran}).
\end{description}

To facilitate the numerics of model B, we have implemented the following procedure. Upon the death of $i$ another individual $j \neq i$ is chosen at random, and is rejected (to reproduce and fill the empty slot with its offspring) with probability $1-\exp(s_j - s_{max})$, where $s_{max}$ is the maximum instantaneous fitness, so if this individual fitness is $s_{max}$ it is accepted with certainty. Upon rejection,  another individual is picked at random and the procedure is iterated until the first acceptation. This can be easily shown to produce the transition probabilities given in Eq. (\ref{moran}).

Our nomenclature is summarized in Table \ref{table1}.

\begin{table}
\begin{center}
\caption {Glossary} \label{table1}
    \begin{tabular}{ | l |  p{10cm} |}

    \hline
    Term  &  Description \\ \hline
    $N$ &  number of individuals in the community. \\ \hline
    $n_i$ &  number of individuals belonging to species $i$. \\ \hline
    $\delta$ & persistence time of the environment, measured in  generations.  \\
    \hline
    $\Pi_{n=1}$ & The chance of a single mutant to establish to reach fixation in a two species game  \\ \hline
     $s$ & the relative (logarithmic) fitness of a species. $s = s_0 \pm \eta(t)$.  \\ \hline
     $s_0$ & the time-independent component of the fitness.   \\ \hline
     $\gamma$ & the amplitude fitness fluctuations.  \\ \hline
    $g \equiv  \delta \gamma^2/2$ & the strength of environmental stochasticity. \\ \hline
    $\tilde{s} \equiv 2 s_0/ \gamma^2 $& variance to mean ratio of selection fluctuations. \\ \hline
    $\nu$ & mutation rate. \\ \hline
    $\Delta s$ & average strength of a single mutation  \\ \hline
    $\overline{s_0}$ & the averaged mean  fitness. Overline implies average over all individuals in the community.  \\ \hline
    $v_{ev}$ & speed of evolution, $d\overline{s_0}/dt$.  \\ \hline
    \end{tabular}

\end{center}

\end{table}

\section{Pure demographic noise} \label{demosection}

In this section, we review shortly some well known results for the speed of evolution in a system with pure demographic noise. In order to set up the framework for our discussion, where we include fluctuating selection. To this end we will try to emphasize two basic aspects of the analysis, namely the chance of establishment of a beneficial mutation and Fisher's fundamental theorem of natural selection.

Let us consider a system in which  $N-1$ individuals have the same fitness $s$ and one beneficial mutant has $s = s + \Delta s$. (for the moment we neglect other mutation events).  The chance of the single mutant to reach fixation is given by~\cite{crow1970introduction}
\begin{equation} \label{dem}
\Pi_{n=1} = \frac{1-e^{-\Delta s}}{1-e^{-N \Delta s}}.
\end{equation}
Throughout this paper we shall assume that $\Delta s \ll 1$.

Eq. (\ref{dem}) has two regimes. In the strong selection regime $N \Delta s \gg 1$, the denominator is essentially unity (if $
\Delta s>0$) and the chance of fixation becomes $N$-independent, so $\Pi_{n=1} \approx \Delta s$.  In the weak selection regime $N \Delta s \ll 1$, the  chance of fixation for a single mutant is given, to first order in $N \Delta s$, by,
\begin{equation} \label{weakN}
\Pi_{n=1} = \frac{1}{N} \left( 1+ \frac{N \Delta s}{2} \right).
\end{equation}

One may  define a critical abundance $n_c = 1/\Delta s$: below this abundance, the mutant population dynamics is dominated by noise and selection is negligible, while above $n_c$ selection is dominant and assures fixation~\cite{desai2007speed}. The weak selection regime corresponds to  $N \ll n_c$, where the dynamics is almost neutral; in a purely neutral system all individuals are demographically equivalent and the chance of fixation is, by symmetry, $1/N$. In the strong selection regime, the mutant abundance has to reach $n_c$ in an almost neutral game to make selection inevitable, and the chance for that is $1/n_c = \Delta s$.

When the mutant is deleterious, $\Delta s <0$, the factor $e^{ -N \Delta s}$ in the denominator of (\ref{dem}) diverges in the strong selection regime, making the chance of fixation exponentially small. For $N \ll n_c$, on the other hand, selection is negligible and the chance of fixation does not change much, as indicated by Eq. (\ref{weakN}).

Now let us analyze the behavior of this system when new mutants are produced at a constant per-generation rate $\nu N$. A (deleterious or advantageous) mutant is ``successful" (will reach fixation) with probability $\Pi_{n=1}(\pm \Delta s)$, so the average number of generations one has to wait until the birth of a successful beneficial mutant is,
\begin{equation}
\tau^+_1 = \frac{2}{\nu N  \Pi_{n=1}(\Delta s)},
\end{equation}
and the mean time until the birth of a successful deleterious mutant is $ \tau^-_1 = 2/[\nu N  \Pi_{n=1}(-\Delta s)]$.

In a model with pure demographic stochasticity, the time to fixation in the weak selection regime is $\tau_2 = N$ generations, while in the strong selection regime~\cite{crow1970introduction},
\begin{equation} \label{tau2}
\tau_{2} = \frac{2 \ln(N)}{|\Delta s|}.
\end{equation}
In both cases this timescale is independent of the sign of $\Delta s$.

Accordingly, when $N \nu$ is small, so that the system is in its successional-fixation (one locus) phase, the speed of evolution is,
\begin{equation} \label{basic}
v_{ev} = \left( \frac{\Delta s}{\tau_1^+ + \tau_2^+}\right)\frac{\Pi^+_{n=1}}{\Pi^+_{n=1}+\Pi^-_{n=1}}+\left(\frac{-\Delta s}{\tau_1^- + \tau_2^-}\right)\frac{\Pi^-_{n=1}}{\Pi^+_{n=1}+\Pi^-_{n=1}},
\end{equation}
where quantities with the superscript plus (minus) are functions of $+\Delta s$ ( $-\Delta s$).  However, unless $\tau_2 \ll \tau_1$, new successful mutants appear before the fixation of the first one occurs. Accordingly, the condition for successional-fixation is

\begin{center}
\begin{tabular}{c c}
%  \hline
   Strong selection \qquad \qquad \qquad&$ \nu N \ln N \ll \Delta s/\Pi^+_{n=1} \approx 1 $  \\
  Weak selection \qquad \qquad \qquad&  $\nu N \ll 1 $ .\\
%  \hline
\end{tabular}
\end{center}

\begin{figure}
\includegraphics[width=9cm]{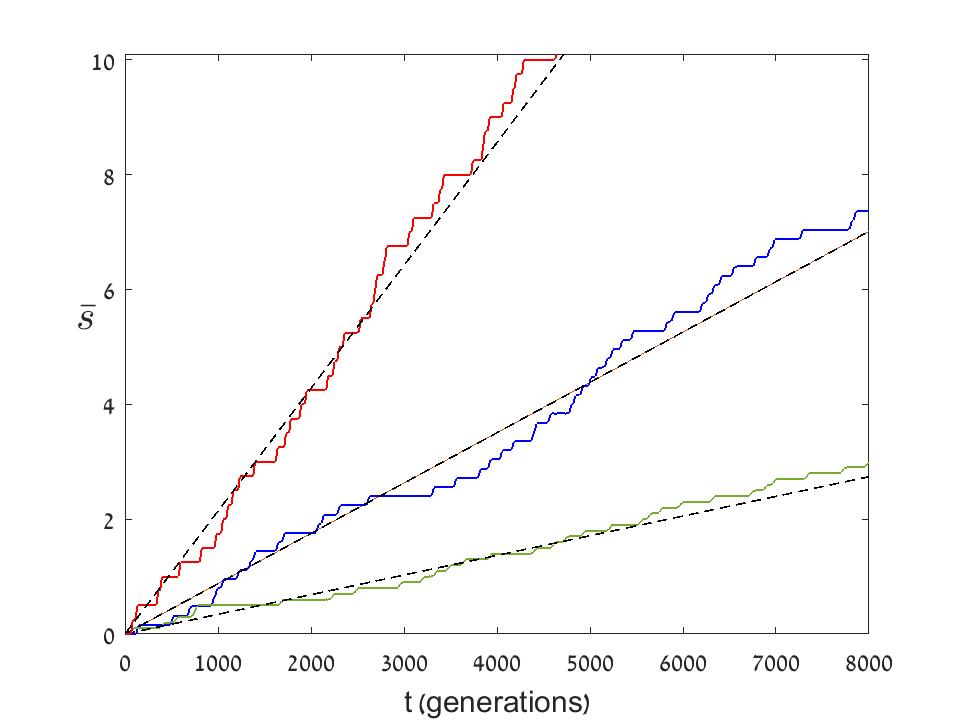}
\caption{The increase in the average fitness of a community, $\bar{s}$, as a function of time (measured in generations). Results are shown  for a community of $N=10^4$ individuals, simulated with $\nu = 10^{-5}$, where $\Delta s = 0.08, \ 0.16$ and $0.25$ (green, blue and res lines, correspondingly).  The dashed black lines are the predictions of (\ref{basic2}) in each case. The pronounced steps in $\bar{s}$ indicate that the system is indeed in its successional-fixation  phase, and the absence of down steps is a manifestation of strong selection.  }\label{demo1}
\end{figure}

In the successional fixation phase Eq. (\ref{basic}) takes the form,
\begin{equation} \label{basic1}
v_{ev} = \frac{N \nu \Delta s }{ 2}\frac{(\Pi^+_{n=1})^2-(\Pi^-_{n=1})^2}{\Pi^+_{n=1}+\Pi^-_{n=1}} = \frac{N \nu \Delta s }{ 2}(\Pi^+_{n=1}-\Pi^-_{n=1}).
\end{equation}
In the strong selection regime $\Pi^-_{n=1}$ vanishes and
\begin{equation} \label{basic2}
v_{ev}   = \frac{N \nu \Delta s  \Pi^+_{n=1}}{ 2} \approx  \frac{N \nu (\Delta s)^2 }{2},
\end{equation}
where the last term comes from Eq. (\ref{dem}) when $\Delta s \ll 1$ and $N \Delta s \gg 1$. This behavior is demonstrated in Figure \ref{demo1}.

In the weak selection regime,
\begin{equation} \label{basic3}
v_{ev} =  N \nu (\Delta s)^2  \frac{d \Pi_{n=1}}{d\Delta s}|_{s=0} = \frac{N \nu (\Delta s)^2 }{2},
\end{equation}
so the speed of evolution is the same in the weak and the strong selection regimes.

When $N$ (or $\nu$) increases the system leaves the successional-fixation phase and enters the clonal interference phase~\cite{gerrish1998fate}, where many clones with different fitness levels compete and mutate at the same time. Unlike \cite{desai2007speed}, here we make no distinction  between ``clonal interference" and ``multiple mutations" as two different approximate pictures of the evolutionary dynamics in the multiple species phase, so our clonal interference regime is everywhere outside the successional-fixation region of parameters.

\begin{figure}
\includegraphics[width=9cm]{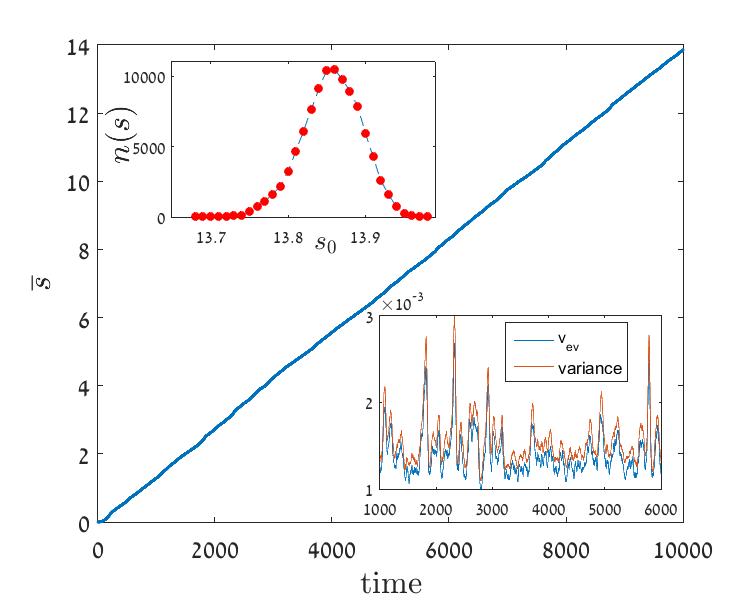}
\caption{The increase in the average fitness of a community, $\bar{s}$, as a function of time (main panel). Results are shown  for a community of $N=10^5$ individuals, simulated with $\nu = 0.1$ and $\Delta s = 0.01$, in the clonal interference phase. A histogram (upper left) shows the number of individuals at each fitness level at the endpoint of the simulation ($t=10000$). Fisher's  fundamental theorem of natural selection is demonstrated in the lower-right inset, where the fitness variance (red) and $v_{ev}$ (local derivative of $\bar{s}(t)$, smoothed over 20 points) are plotted together.    }\label{demo2}
\end{figure}

In the clonal interference phase the growth of $\bar{s}$ is continuous, not stepwise, since new beneficial mutations appear (and the abundance of  their  lineage grow) in parallel. Now the steady state is a soliton that moves, on average, in a constant speed. Since the growth rate of the most beneficial clone (at the leading edge of the soliton) is determined by its fitness advantage with respect to the average individual, it is clear that the width of the soliton sets the speed of evolution, in agreement with  Fisher's  fundamental theorem of natural selection.   Figure \ref{demo2} demonstrates this aspect of the theory using our simulations.

\section{Speed of evolution in the presence of environmental stochasticity}

The fate of a mutant population in a  two-species community has been studied recently~\cite{danino2016stability,danino2017fixation,meyer2018noise}. The chance of fixation and the time to absorption (either fixation or loss) were calculated for both model A and model B. The mean time to fixation was calculated only for model A. In this section we will use these results.

As explained above, the environmental noise is characterized by its amplitude, which is proportional to $\gamma$, and its persistence time $\delta$. The strength of the environmental noise is $g = \gamma^2 \delta /2$, and the ratio between this strength and the strength of demographic fluctuations, $1/N$, is given by $G \equiv Ng$. Another measure is the ratio between selection and environmental stochasticity, $\tilde{s} \equiv 2 |\Delta s|/g$ (see Table \ref{table1}).

An important feature of both model A and model B is the appearance of a new scale that separates the fluctuation dominated regime from the selection dominated regime. This new scale is,
\begin{equation}
n_c = \frac{\exp \left( \frac{g}{2|\Delta s|} \right)}{g},
\end{equation}
 and it diverges when $|\Delta s| \ll g$. Accordingly, the study of the weak selection regime  is much more relevant when a system with environmental stochasticity is considered.

Note that in \cite{cvijovic2015fate} a very similar scale,  $(\exp(g/\Delta s)-1)/g$, was introduced. This definition has the advantage that the critical abundance converges to $1/s$ when $g$ goes to zero, but otherwise the two scales are qualitatively similar. In any case, the results presented below were obtained using an asymptotic analysis in the large $G=Ng$ limit, so we do not expect them to converge to the purely demographic results when $g \to 0$.

\subsection{Model A}

Both the chance of fixation and the time to fixation were calculated, for model A, in our recent work~\cite{danino2017fixation}. The chance of fixation of a single mutant is,
\begin{equation} \label{1A}
\Pi_{n=1}  = \frac{1-(1+g)^{-(\Delta s/g)}}{1-G^{-2 \Delta s/g}},
\end{equation}
 Eq. (\ref{1A}) yields, in the weak selection regime~\cite{cvijovic2015fate,danino2017fixation},
\begin{equation} \label{weakA}
\Pi_{n=1} \approx \frac{\ln(1+g)}{2\ln(G)}\left(1+ \frac{\Delta s}{g}\ln G \right).
\end{equation}

The time to fixation in the strong selection regime is the same as  in the purely demographic case (\ref{tau2}), but in the weak selection sector~\cite{danino2017fixation},
\begin{equation} \label{eq210}
\tau_2 = \frac{2}{3g} \ln^2(gN).
\end{equation}

Accordingly, the system will be in its successional-fixation phase if,

\begin{center}
\begin{tabular}{c c}
%  \hline
   Strong selection \qquad \qquad \qquad&$ \nu N \ln N \ll \frac{\Delta s}{1-(1+g)^{-\Delta s/g}} \approx 1$  \\
  Weak selection \qquad \qquad \qquad&  $\nu N \ln (Ng) \ll \frac{6g}{\ln(1+g)} \approx 6 $ .\\
%  \hline
\end{tabular}
\end{center}

As in the pure demographic case, the speed of evolution in the weak and the strong selection regimes turns out to be the same,
\begin{equation} \label{velA}
v_{ev} =  \frac{N \nu (\Delta s)^2 }{2}\frac{\ln(1+g)}{g}.
\end{equation}

Since $\ln(1+g)/g$ is a monotonously decreasing function of $g$, the speed of evolution in model A decreases, in comparison with its pure demographic value, when the environmental stochasticity increases.  Taking $\delta = 3$ and $\gamma = \sqrt{2}$, we were able to compare our prediction  with the numerics in the regime where the outcome of (\ref{velA}) differs substantially from the demographic noise prediction (\ref{basic3}). The results are shown in Figure \ref{maritan}. The agreement between  (\ref{velA}) and the outcome of the simulation is evident, as well as the disagreement between the numerics and (\ref{basic3}). Moreover, one can see that the stochasticity actually slows down the speed of evolution, as expected.

\begin{figure*}
\includegraphics[width=8cm]{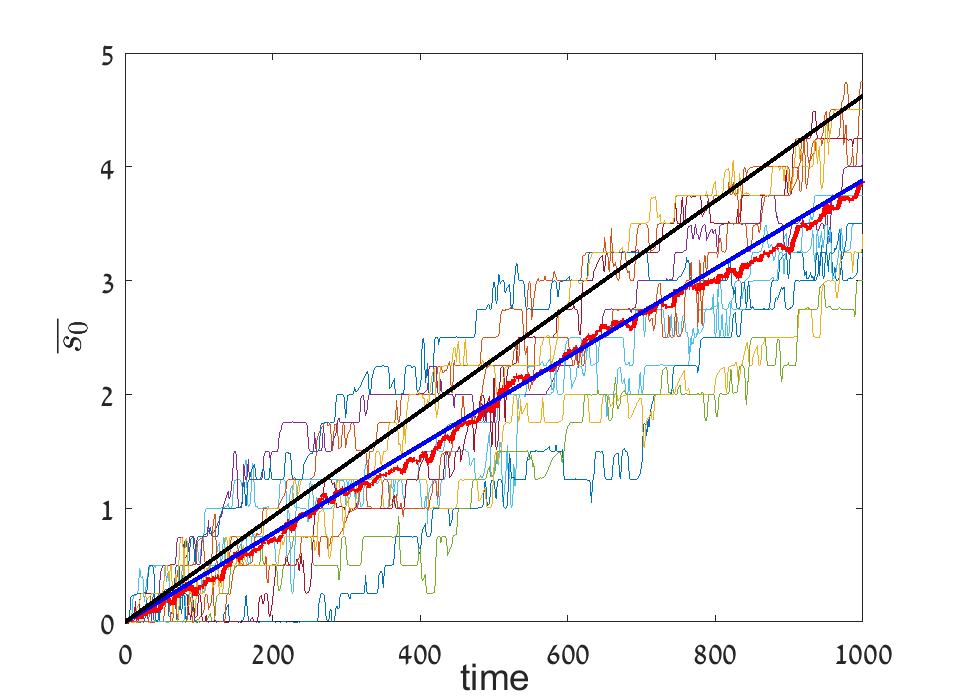}
\includegraphics[width=8cm]{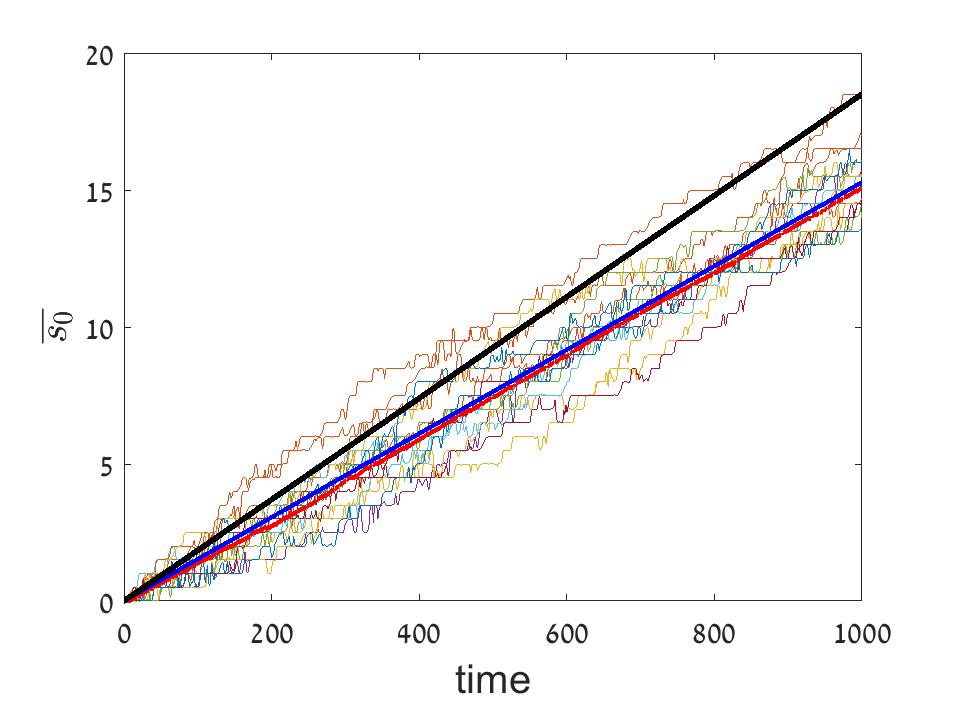}
\caption{$\bar{s_0}$ vs. time in the successional-fixation phase of model A. Results are shown  for $N=10^5$, $\gamma = \sqrt{2}$, $\delta =3$, with $\Delta s = 0.25$ (left panel) and  $\Delta s = 0.5$ (right panel). Each run yields a single curve and one can see the pronounced stepwise structure. The average (thick red line) is very close to the prediction of (\ref{velA}) (thick blue curve) and differs substantially from the demographic noise prediction (\ref{basic3}) (thick black line). In the left panel the average was taken from 10 runs. In the right panel the average was taken from 100 runs, only a few of them are shown.   }\label{maritan}
\end{figure*}

\begin{figure}
\includegraphics[width=9cm]{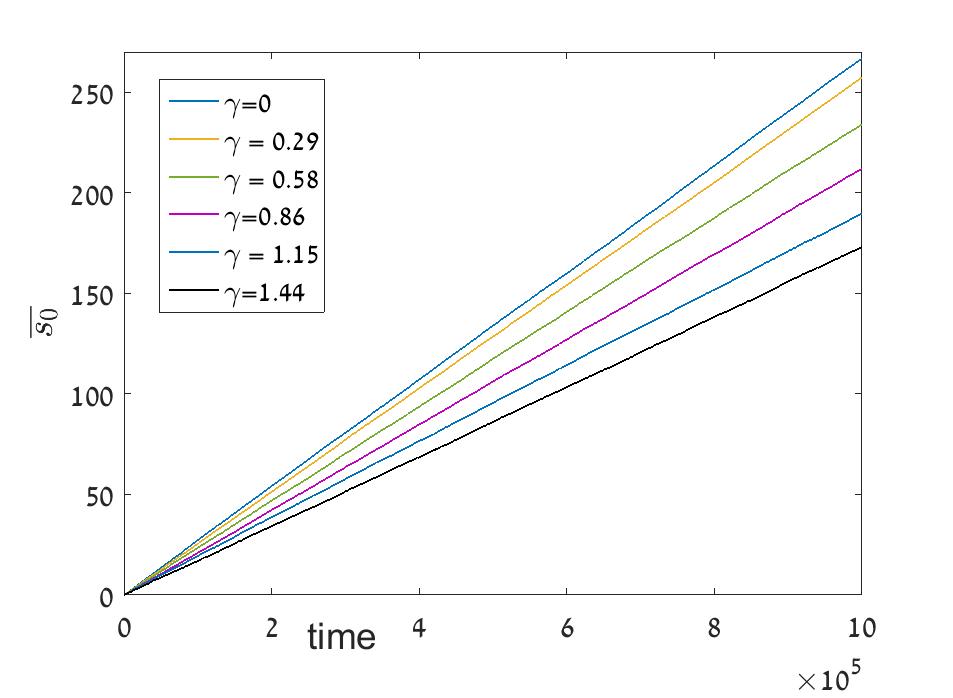}
\caption{The increase in the average fitness of a community as a function of time. Results are shown here for a community of $N=10^4$ individuals, simulated with $\delta = 0.1$, $\nu = 0.01$ and $\delta s =0.01$. The  lines show the average fitness $\overline{s_0}$ as a function of time, where time is measured in generations. The speed of evolution is highest when $\gamma = 0$ (blue line) and decreases as $\gamma$ grows. }\label{fig1}
\end{figure}

\begin{figure}
\includegraphics[width=9cm]{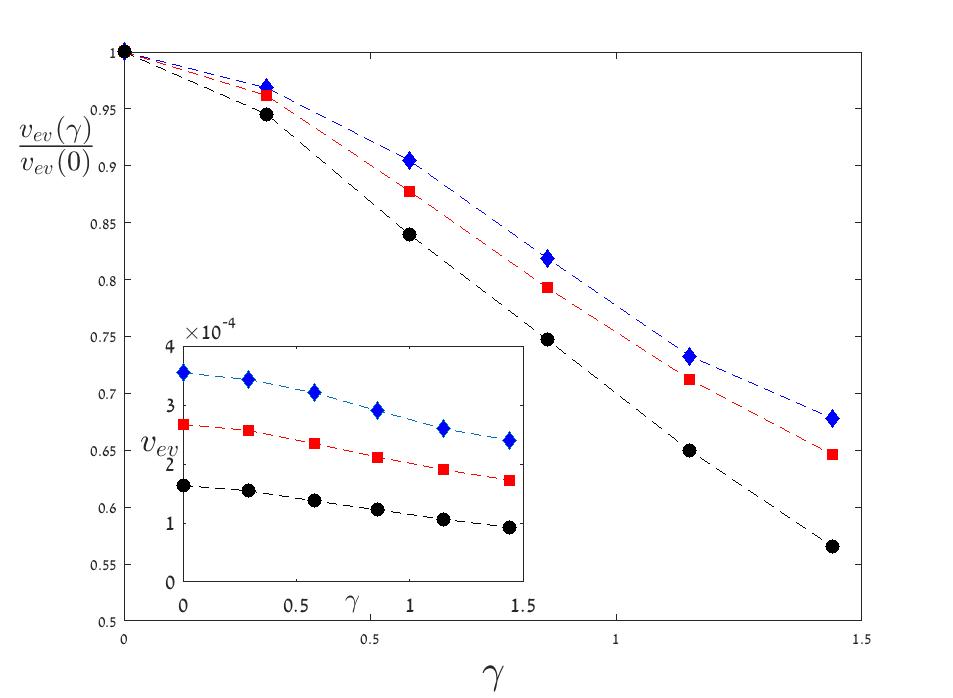}
\caption{The speed of evolution  for model A (the slope measured from curves like those presented in Fig. \ref{fig1}) when plotted against $\gamma$ for $N=1000$ (black full line and circles), $N=10^4$ (red dashed line and squares) and $N=10^5$  (blue dotted line and diamonds). All slopes were obtained from simulation of $10^6$ generations, with $\Delta s = 0.01$  and $\nu = 0.01$. The speed of evolution increases with $N$ but for each $N$ it decreases with $\gamma$ (inset). The relative decrease with $\gamma$, $v_{ev}(\gamma)/v_{ev}(\gamma=0)$ becomes slightly smaller as $N$ increases (main panel).    }\label{fig2}
\end{figure}

\begin{figure}
\includegraphics[width=9cm]{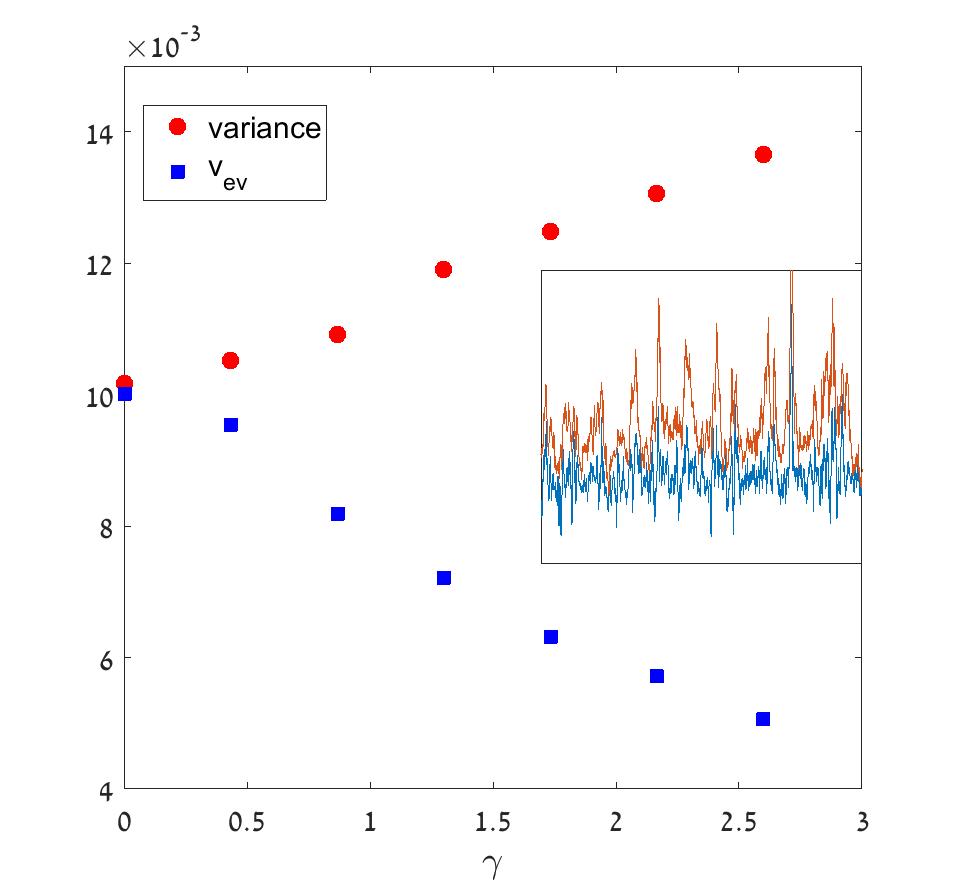}
\caption{The speed of evolution   (the slope measured from curves like those presented in Fig. \ref{fig1}, blue squares) and the average variance of $s_0$ (red circles), both plotted against $\gamma$ for for simulations with  $N=10^4$, $\nu = 0.01$, $\delta = 0.1$ and $\Delta s=0.1$. For each value of $\gamma$ the data were extracted from a run with $10^4$ generations. In the inset one can see the instantaneous values of the variance of $s_0$ and the velocity as measured during 1000 generations of the run with $\gamma = 2.6$, where the differences between the two quantities are pronounced.     }\label{fig3}
\end{figure}

In the clonal interference phase we have no analytic predictions and the main  numerical results are illustrated in Figures \ref{fig1}, \ref{fig2}, and \ref{fig3}. The mean fitness of the whole community still grows linearly in time, and the speed of evolution decreases monotonically as  $\gamma$ increases in agreement with the outcome in the successional-fixation phase. For fixed $\gamma$, the relative effect of the environmental noise, $v_{ev}(\gamma)/v_{ev}(\gamma=0)$, decreases as $N$ grows, as shown in Figure \ref{fig2}, but the rate of decrease slows down considerably with $N$.  Although the relative effect becomes smaller as $N$ increases, it is still pronounced. Since the speed of evolution diverges as $N \to \infty$ it is difficult to speak about the effect of environmental noise in the  asymptotic limit.

Finally, let us consider Fisher's fundamental theorem of natural selection under environmental stochasticity.  While at each moment the rate of growth of the  instantaneous fitness of the community, $d\overline{s(t)}/dt$, is equal to the instantaneous variance of $s$, the same statement is not true with regard to $s_0$. As demonstrated in Figure \ref{fig3}, as $\gamma$ grows the  variance of $s_0$  (when averaged over long periods of time) \emph{increases} while the values of $d\overline{s_0}/dt$ decrease.

\subsection{Model B}

The noise-induced stabilizing mechanism that affects model B dynamics has been analyzed in detail in \cite{danino2016effect,danino2016stability,meyer2018noise}. For a two-species competition, where the fitness difference is $\Delta s$, the noise-induced attractive fixed point is at
\begin{equation} \label{fp}
n^* = N \left( \frac{1}{2} + \frac{\Delta \tilde{s}}{2} \right).
\end{equation}
where $ \Delta \tilde{s} = 2 \Delta s/\gamma^2$.  When $n^*/N$ is not between zero and one (i.e. $|\Delta \tilde{s}|>1$), the qualitative behavior of model B is similar to that of model A~\cite{meyer2018noise} and we will not discuss it here. Instead we are looking at cases where the attractive fixed point is in the physical domain. For this scenario  the chance of fixation of a single mutant was calculated in ~\cite{meyer2018noise},
\begin{equation} \label{1B}
\Pi_{n=1}  = \frac{1-(1+g)^{-(1+\Delta \tilde{s})/\delta}}{1+D_1 G^{-2 \Delta \tilde{s}/\delta}},
\end{equation}
where $D_1= (1+\Delta \tilde{s})/(1-\Delta \tilde{s})$.

Unlike model A, here in the weak selection regime the chance of fixation approaches an $N$-independent constant,
\begin{equation} \label{weakB}
\Pi_{n=1} \approx \frac{1-(1+g)^{-1/\delta}}{2} \left(1 + \frac{ \Delta \tilde{s} \ln(G)}{\delta}\right).
\end{equation}

The time to fixation $\tau_2$ has not yet calculated  for model B. However, in \cite{hidalgo2017species,danino2016stability}  the time to absorption (either fixation or extinction) for a mutant population that reaches $n^*$ was found,
\begin{equation}
\tau_{abs} \sim C_1 N^{\frac{1-\Delta \tilde{s}}{\delta}}
\end{equation}
where $C_1$ is an $N$-independent constant.

In \cite{meyer2018noise} we showed that the numerator of (\ref{1B}) is the chance of the mutant to reach the basin of attraction of the coexistence fixed point (\ref{fp}), while the denominator is one over the chance to reach fixation starting from the noise-induced fixed point. We also showed that the distance that a mutant should travel in order to reach the basin of attraction, and the time (in generation) required for that, are both $N$-independent. Accordingly, the leading contribution to $\tau_2$ in the large $N$ limit comes from $\tau_{abs}$, and for the rest of this subsection we take $\tau_2 \approx \tau_{abs} = C_1 N^{(1-\Delta \tilde{s})/\delta}$.

The divergence of $\tau_2$ implies that the successional-fixation phase is extremely narrow under model B dynamics. For model B the both $\Pi_{n=1}$ and $\tau_2$ change only slightly between the weak and the strong selection regimes, so the condition for the successional-fixation phase is
\begin{equation}
\nu N^{1+\frac{1-\Delta \tilde{s}}{\delta}} \ll C_2,
\end{equation}
where $C_2$ is, again, an $N$-independent constant.

\begin{figure}
\includegraphics[width=9cm]{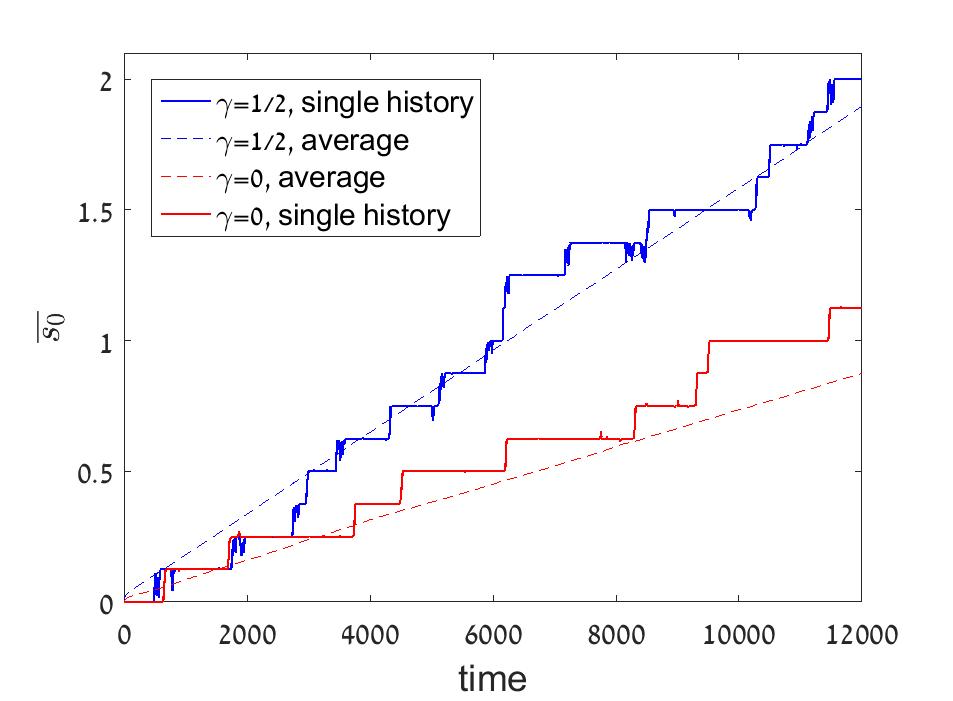}
\caption{$\overline{s_0}$ vs. time at the successional-fixation phase for the purely demographic case $\gamma =0$  and for a system in which environmental stochasticity was implemented according to the procedure of model B. A single typical history (solid line) and an average over 3000 samples (dashed lines) are shown for $\gamma=0$ (red) and $\gamma=1/2$ (blue), with $\delta = 0.2$, $\nu = 10^{-4}$, $N=100$ and $ \Delta s = 0.125$. In both cases the system is in its strong selection regime. The speed of evolution in the neutral case is, for these parameters, $1/12800$, while our expression for strong selection model B dynamics yields, for $\gamma =1/2$, $v_{ev} \approx 1/7300$, both estimations fit quite well the observed results. Clearly the speed of evolution is growing with $\gamma$, as opposed to the  behavior without storage (model A) demonstrated  in Figure \ref{maritan}.   }\label{chesson_staircase}
\end{figure}

\begin{figure}
\includegraphics[width=9cm]{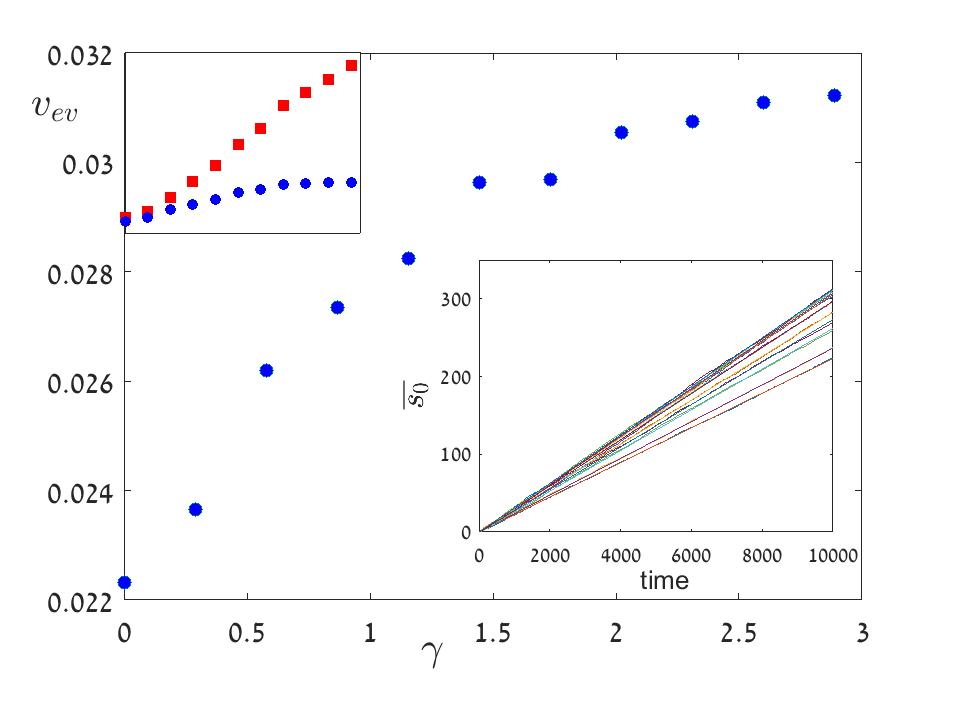}
\caption{$v_{ev}$ vs. $\gamma$ in the clonal interference phase of model B. For 11 values of $\gamma$, $\overline{s_0}$ was plotted against time for 10000 generations (lower inset) and the measured slope was plotted against  $\gamma$ (main, blue circles). As in the successional-fixation phase, $v_{ev}$ increases with $\gamma$.  In the upper left inset the same  $v_{ev}$ points (blue) are shown together with the variance of $s_0$ (red): while $v_{ev}$ grows, the variance grow even faster.     The parameters of the simulation were $N = 1000$, $\nu =0.01$, $\delta = 0.1$ and $\Delta s=0.2$.    }\label{chesson2}
\end{figure}

The speed of evolution in these two regimes is,

\begin{center}
\begin{tabular}{c c}
%  \hline
   Strong selection \qquad \qquad \qquad&$ v_{ev} = \frac{\nu N \Delta s}{2} [1-(1+g)^{-1/\delta - \Delta s/g}] $  \\
  Weak selection \qquad \qquad \qquad&  $ v_{ev} = \frac{\nu N \ln (Ng) (\Delta s)^2}{2g} [1-(1+g)^{-1/\delta}]  $ .\\
%  \hline
\end{tabular}
\end{center}
Note that these two expressions coincide where $N \sim n_c$. In contrast with model A, here the speed of evolution increases with $\gamma$, the amplitude of the environmental fluctuations, as demonstrated in Figure  (\ref{chesson_staircase}). In the clonal interference phase we observe numerically the same behavior: the velocity grows with $\gamma$, first almost linearly and then the graph levels off (Figure \ref{chesson2}). In the region of parameters we have checked, our simulations (not shown) also suggest that the relative increase in the velocity becomes larger as $N$ increases, as opposed to the outcome of the same numerical experiment in model A.

Again, we observe  that  Fisher's fundamental theorem does not hold for the (time averaged) variance of the time independent component of the fitness $s_0$, since it  grows even faster than  $v_{ev}$ (inset of  Figure \ref{chesson2}).

\section{Discussion}

The ``speed of evolution" problem  belongs to the general field of spatial invasion of a stable state into an unstable one. Not surprisingly, the first to consider this problem was R. A. Fisher \cite{fisher1937wave}, who suggested  his famous equation,
\begin{equation} \label{fisher}
\frac{\partial \rho(x,t)}{ \partial t} = D \frac{\partial^2 \rho(x,t)}{\partial x^2} + a [\rho(x,t) - \rho^2(x,t)],
\end{equation}
to describe the spreading of a favored mutation in a spatially structured population. Here $\rho(x,t)$ is the fraction of the local population that have the favored mutation, $a$ is the local growth rate of this fraction (it is proportional to the selective advantage of the mutation) and $D$ is the spatial diffusion constant. The Fisher (FKPP) equation is known to support front that propagates with velocity $2 \sqrt{Da}$; this velocity is determined, as in other cases of what is known as ``pulled fronts",  by the dynamics of its leading edge \cite{van2003front}.

In the context of evolution, space is translated to fitness ($x \to s$), the diffusion constant is  the effective mutation rate and $a$, the growth rate of a clone with fitness $s$, is simply $s-\bar{s}$, the distance from the mean.  The corresponding equation is,
\begin{equation} \label{fl}
\frac{\partial \rho(s,t)}{ \partial t} = \nu_{eff} \frac{\partial^2 \rho(s,t)}{\partial x^2} + \rho(s,t) \int_{-\infty}^{\infty} \ ds' \  (s-s')\rho(s',t)  = \nu_{eff} \frac{\partial^2 \rho(s,t)}{\partial x^2} + (s-\bar{s}) \rho(s,t).
\end{equation}
This is  a Fisher-like equation with nonlocal competition, but the competition kernel is asymmetric and increases with the distance, unlike the cases that were studied in the literature in this context \cite{maruvka2006nonlocal,fuentes2003nonlocal}. Multiplying  both sides of  Eq. (\ref{fl}) by $s$ and integrating over $s$ one may easily derive the relationship $\dot{\bar{s}} = \overline{s^2} - \bar{s}^2$, i.e.,  Fisher's fundamental theorem of natural selection.

It may be very interesting to consider the effect of demographic and environmental stochasticity on the features of (\ref{fl}) by adding terms like $\zeta(x,t) \sqrt{\rho}$ (for demographic stochasticity) and $\xi(x,t) \rho $ (for environmental stochasticity), where $\zeta$ and $\xi$ are white noise. The resulting equations may be analyzed perturbatively  to reveal the effect of noise in a neatly arranged fashion.

In the strong selection regime we have seen that the effect of stochasticity on the successful mutation rate is the crucial factor that determines the speed of evolution in the successional-fixation phase: $v_{ev}$ grows when $\Pi_{n=1}$ increases, as in model B, and slows down when $\Pi_{n=1}$ decreases for model A. In the clonal interference phase our simulations suggest the same behavior, despite the fact that the width of the soliton (the variance of $s_0$)  grows when $g$ increases in both cases. Apparently the width of the soliton plays a secondary role with respect to the successful mutation rate. Given that, we expect that the main effect of the addition of noise to (\ref{fl}) will be  a renormalization of the ``diffusion constant" $\nu$, which in turn affects the width of the  moving soliton.

In the weak selection regime the interpretation of our results is more subtle. Here  both deleterious and beneficial mutants may reach fixation, and the speed of evolution depends on the linear response of $\Pi$ to $\Delta s$. In model A environmental variations act to diminish the differences between positive and negative selection, hence the speed of evolution decreases when the fluctuations increases. In model B the main effect of environmental variations is to stabilize the attractive fixed point at $n^*$, and this point moves slightly to the right (left) of $N/2$ when $\Delta s$ is positive (negative), yielding  a larger response than in the purely demographic case.

The validity of our analysis is determined by the validity of the underlying assumption that lead to the expressions presented here for the chance of (and the time to) fixation. In particular, our theory is based on a large $G$  asymptotic analysis and it usually fails when $G \sim 1$, so our result do not extrapolate, in general, to the purely neutral limit (although some of them do). Moreover, our methods assume that environmental noise generates diffusive motion in the log-abundance space, so if $\delta \gg  \ln N/\Delta s$, so  that fixation may occur during a single sweep, the results become invalid.

The relevance of this work depends on the importance of fitness fluctuations in the evolutionary process. Empirical studies of abundance variations in a wide variety of ecosystems, from animals and trees to microbial experiments, show decisively the dominance of environmental stochasticity~\cite{hekstra2012contingency,chisholm2014temporal,kalyuzhny2014niche,
kalyuzhny2014temporal,fung2016reproducing}. Much less is known about the the role of environmental variations during evolution, and even empirical assessments of fitness diversity in a population are quite rare  \cite{saether2015concept}.  

In a recent review article~\cite{messer2016can}, Messer et al. discussed some evidence for the effect of environmental variations on the magnitude and the sign of selective forces and their possible effect on the evolutionary dynamics. For example, measured rates of phenotypic evolution are inversely proportional to the timescale of measurement, a scaling that arises naturally if the population tracks environmental variations by heritable changes in   phenotypic traits.  In general, we see no a-priori reason to believe that in a generic evolutionary process $\Delta s$ is much larger than $\gamma^2$. If this is not the case, our analysis may be important for the study of the evolutionary process in large and even to the eco-evolutionary dynamics observed in microbial communities.

In a natural community that evolves via selection and mutation, the response of different species to environmental variations is perhaps partially correlated. In a two species game, for example, environmental variations that affect coherently both the mother and the daughter species will change the overall carrying capacity $N$. This will modify the effective strength of the demographic noise keeping all other parameters fixed, a scenario that was  discussed recently by Wienand et al.~\cite{wienand2017evolution}.
 
In parallel, the strength of competition between species may decay with the distance to their common ancestor (the competitive-relatedness hypothesis), but the typical fitness differences will grow. The relationship between these two factors has been analyzed in~\cite{shtilerman2015emergence}. A reliable entanglement of these effects in an empirical system may prove a formidable task, but the basic intuitive argument presented in this work can provide a few simple guidelines for such an analysis.

{\bf Acknowledgments} N.M.S. acknowledge the support of the ISF-NRF Singapore joint research program (grant number 2669/17).

\bibliography{refs_JSP}

\begin{thebibliography}{42}
\expandafter\ifx\csname natexlab\endcsname\relax\def\natexlab#1{#1}\fi
\expandafter\ifx\csname bibnamefont\endcsname\relax
  \def\bibnamefont#1{#1}\fi
\expandafter\ifx\csname bibfnamefont\endcsname\relax
  \def\bibfnamefont#1{#1}\fi
\expandafter\ifx\csname citenamefont\endcsname\relax
  \def\citenamefont#1{#1}\fi
\expandafter\ifx\csname url\endcsname\relax
  \def\url#1{\texttt{#1}}\fi
\expandafter\ifx\csname urlprefix\endcsname\relax\def\urlprefix{URL }\fi
\providecommand{\bibinfo}[2]{#2}
\providecommand{\eprint}[2][]{\url{#2}}

\bibitem[{\citenamefont{Tsimring et~al.}(1996)\citenamefont{Tsimring, Levine,
  and Kessler}}]{tsimring1996rna}
\bibinfo{author}{\bibfnamefont{L.~S.} \bibnamefont{Tsimring}},
  \bibinfo{author}{\bibfnamefont{H.}~\bibnamefont{Levine}}, \bibnamefont{and}
  \bibinfo{author}{\bibfnamefont{D.~A.} \bibnamefont{Kessler}},
  \bibinfo{journal}{Physical review letters} \textbf{\bibinfo{volume}{76}},
  \bibinfo{pages}{4440} (\bibinfo{year}{1996}).

\bibitem[{\citenamefont{Rouzine et~al.}(2003)\citenamefont{Rouzine, Wakeley,
  and Coffin}}]{rouzine2003solitary}
\bibinfo{author}{\bibfnamefont{I.~M.} \bibnamefont{Rouzine}},
  \bibinfo{author}{\bibfnamefont{J.}~\bibnamefont{Wakeley}}, \bibnamefont{and}
  \bibinfo{author}{\bibfnamefont{J.~M.} \bibnamefont{Coffin}},
  \bibinfo{journal}{Proceedings of the National Academy of Sciences}
  \textbf{\bibinfo{volume}{100}}, \bibinfo{pages}{587} (\bibinfo{year}{2003}).

\bibitem[{\citenamefont{Park and Krug}(2007)}]{park2007clonal}
\bibinfo{author}{\bibfnamefont{S.-C.} \bibnamefont{Park}} \bibnamefont{and}
  \bibinfo{author}{\bibfnamefont{J.}~\bibnamefont{Krug}},
  \bibinfo{journal}{Proceedings of the National Academy of Sciences}
  \textbf{\bibinfo{volume}{104}}, \bibinfo{pages}{18135}
  (\bibinfo{year}{2007}).

\bibitem[{\citenamefont{Desai et~al.}(2007)\citenamefont{Desai, Fisher, and
  Murray}}]{desai2007speed}
\bibinfo{author}{\bibfnamefont{M.~M.} \bibnamefont{Desai}},
  \bibinfo{author}{\bibfnamefont{D.~S.} \bibnamefont{Fisher}},
  \bibnamefont{and} \bibinfo{author}{\bibfnamefont{A.~W.}
  \bibnamefont{Murray}}, \bibinfo{journal}{Current biology}
  \textbf{\bibinfo{volume}{17}}, \bibinfo{pages}{385} (\bibinfo{year}{2007}).

\bibitem[{\citenamefont{Bell}(2010)}]{bell2010fluctuating}
\bibinfo{author}{\bibfnamefont{G.}~\bibnamefont{Bell}},
  \bibinfo{journal}{Philosophical Transactions of the Royal Society of London
  B: Biological Sciences} \textbf{\bibinfo{volume}{365}}, \bibinfo{pages}{87}
  (\bibinfo{year}{2010}).

\bibitem[{\citenamefont{Bergland et~al.}(2014)\citenamefont{Bergland, Behrman,
  O'Brien, Schmidt, and Petrov}}]{bergland2014genomic}
\bibinfo{author}{\bibfnamefont{A.~O.} \bibnamefont{Bergland}},
  \bibinfo{author}{\bibfnamefont{E.~L.} \bibnamefont{Behrman}},
  \bibinfo{author}{\bibfnamefont{K.~R.} \bibnamefont{O'Brien}},
  \bibinfo{author}{\bibfnamefont{P.~S.} \bibnamefont{Schmidt}},
  \bibnamefont{and} \bibinfo{author}{\bibfnamefont{D.~A.}
  \bibnamefont{Petrov}}, \bibinfo{journal}{PLoS Genetics}
  \textbf{\bibinfo{volume}{10}}, \bibinfo{pages}{e1004775}
  (\bibinfo{year}{2014}).

\bibitem[{\citenamefont{Gavrilets}(2010)}]{gavrilets2010high}
\bibinfo{author}{\bibfnamefont{S.}~\bibnamefont{Gavrilets}},
  \bibinfo{journal}{Evolution: the extended synthesis (eds M Pigliucci, GB
  M{\"u}ller)} pp. \bibinfo{pages}{45--79} (\bibinfo{year}{2010}).

\bibitem[{\citenamefont{Messer et~al.}(2016)\citenamefont{Messer, Ellner, and
  Hairston}}]{messer2016can}
\bibinfo{author}{\bibfnamefont{P.~W.} \bibnamefont{Messer}},
  \bibinfo{author}{\bibfnamefont{S.~P.} \bibnamefont{Ellner}},
  \bibnamefont{and} \bibinfo{author}{\bibfnamefont{N.~G.}
  \bibnamefont{Hairston}}, \bibinfo{journal}{Trends in Genetics}
  \textbf{\bibinfo{volume}{32}}, \bibinfo{pages}{408} (\bibinfo{year}{2016}).

\bibitem[{\citenamefont{Dean et~al.}(2017)\citenamefont{Dean, Lehman, and
  Yi}}]{dean2017fluctuating}
\bibinfo{author}{\bibfnamefont{A.~M.} \bibnamefont{Dean}},
  \bibinfo{author}{\bibfnamefont{C.}~\bibnamefont{Lehman}}, \bibnamefont{and}
  \bibinfo{author}{\bibfnamefont{X.}~\bibnamefont{Yi}},
  \bibinfo{journal}{Genetics} \textbf{\bibinfo{volume}{205}},
  \bibinfo{pages}{1271} (\bibinfo{year}{2017}).

\bibitem[{\citenamefont{Steinberg and
  Ostermeier}(2016)}]{steinberg2016environmental}
\bibinfo{author}{\bibfnamefont{B.}~\bibnamefont{Steinberg}} \bibnamefont{and}
  \bibinfo{author}{\bibfnamefont{M.}~\bibnamefont{Ostermeier}},
  \bibinfo{journal}{Science advances} \textbf{\bibinfo{volume}{2}},
  \bibinfo{pages}{e1500921} (\bibinfo{year}{2016}).

\bibitem[{\citenamefont{Taute et~al.}(2014)\citenamefont{Taute, Gude, Nghe, and
  Tans}}]{taute2014evolutionary}
\bibinfo{author}{\bibfnamefont{K.~M.} \bibnamefont{Taute}},
  \bibinfo{author}{\bibfnamefont{S.}~\bibnamefont{Gude}},
  \bibinfo{author}{\bibfnamefont{P.}~\bibnamefont{Nghe}}, \bibnamefont{and}
  \bibinfo{author}{\bibfnamefont{S.~J.} \bibnamefont{Tans}},
  \bibinfo{journal}{Trends in Genetics} \textbf{\bibinfo{volume}{30}},
  \bibinfo{pages}{192} (\bibinfo{year}{2014}).

\bibitem[{\citenamefont{Engen and S{\ae}ther}(2014)}]{engen2014evolution}
\bibinfo{author}{\bibfnamefont{S.}~\bibnamefont{Engen}} \bibnamefont{and}
  \bibinfo{author}{\bibfnamefont{B.-E.} \bibnamefont{S{\ae}ther}},
  \bibinfo{journal}{Evolution} \textbf{\bibinfo{volume}{68}},
  \bibinfo{pages}{854} (\bibinfo{year}{2014}).

\bibitem[{\citenamefont{S{\ae}ther and Engen}(2015)}]{saether2015concept}
\bibinfo{author}{\bibfnamefont{B.-E.} \bibnamefont{S{\ae}ther}}
  \bibnamefont{and} \bibinfo{author}{\bibfnamefont{S.}~\bibnamefont{Engen}},
  \bibinfo{journal}{Trends in ecology \& evolution}
  \textbf{\bibinfo{volume}{30}}, \bibinfo{pages}{273} (\bibinfo{year}{2015}).

\bibitem[{\citenamefont{Cvijovi{\'c} et~al.}(2015)\citenamefont{Cvijovi{\'c},
  Good, Jerison, and Desai}}]{cvijovic2015fate}
\bibinfo{author}{\bibfnamefont{I.}~\bibnamefont{Cvijovi{\'c}}},
  \bibinfo{author}{\bibfnamefont{B.~H.} \bibnamefont{Good}},
  \bibinfo{author}{\bibfnamefont{E.~R.} \bibnamefont{Jerison}},
  \bibnamefont{and} \bibinfo{author}{\bibfnamefont{M.~M.} \bibnamefont{Desai}},
  \bibinfo{journal}{Proceedings of the National Academy of Sciences}
  \textbf{\bibinfo{volume}{112}}, \bibinfo{pages}{E5021}
  (\bibinfo{year}{2015}).

\bibitem[{\citenamefont{Lande et~al.}(2003)\citenamefont{Lande, Engen, and
  Saether}}]{lande2003stochastic}
\bibinfo{author}{\bibfnamefont{R.}~\bibnamefont{Lande}},
  \bibinfo{author}{\bibfnamefont{S.}~\bibnamefont{Engen}}, \bibnamefont{and}
  \bibinfo{author}{\bibfnamefont{B.-E.} \bibnamefont{Saether}},
  \emph{\bibinfo{title}{Stochastic population dynamics in ecology and
  conservation}} (\bibinfo{publisher}{Oxford University Press},
  \bibinfo{year}{2003}).

\bibitem[{\citenamefont{Chesson and Warner}(1981)}]{chesson1981environmental}
\bibinfo{author}{\bibfnamefont{P.~L.} \bibnamefont{Chesson}} \bibnamefont{and}
  \bibinfo{author}{\bibfnamefont{R.~R.} \bibnamefont{Warner}},
  \bibinfo{journal}{American Naturalist} pp. \bibinfo{pages}{923--943}
  (\bibinfo{year}{1981}).

\bibitem[{\citenamefont{Hatfield and Chesson}(1989)}]{hatfield1989diffusion}
\bibinfo{author}{\bibfnamefont{J.~S.} \bibnamefont{Hatfield}} \bibnamefont{and}
  \bibinfo{author}{\bibfnamefont{P.~L.} \bibnamefont{Chesson}},
  \bibinfo{journal}{Theoretical Population Biology}
  \textbf{\bibinfo{volume}{36}}, \bibinfo{pages}{251} (\bibinfo{year}{1989}).

\bibitem[{\citenamefont{Chesson}(1994)}]{chesson1994multispecies}
\bibinfo{author}{\bibfnamefont{P.}~\bibnamefont{Chesson}},
  \bibinfo{journal}{Theoretical Population Biology}
  \textbf{\bibinfo{volume}{45}}, \bibinfo{pages}{227} (\bibinfo{year}{1994}).

\bibitem[{\citenamefont{Fisher}(1930)}]{fisher1930genetical}
\bibinfo{author}{\bibfnamefont{R.~A.} \bibnamefont{Fisher}},
  \emph{\bibinfo{title}{The genetical theory of natural selection: a complete
  variorum edition}} (\bibinfo{publisher}{Oxford University Press},
  \bibinfo{year}{1930}).

\bibitem[{\citenamefont{Frank and Slatkin}(1992)}]{frank1992fisher}
\bibinfo{author}{\bibfnamefont{S.~A.} \bibnamefont{Frank}} \bibnamefont{and}
  \bibinfo{author}{\bibfnamefont{M.}~\bibnamefont{Slatkin}},
  \bibinfo{journal}{Trends in Ecology \& Evolution}
  \textbf{\bibinfo{volume}{7}}, \bibinfo{pages}{92} (\bibinfo{year}{1992}).

\bibitem[{\citenamefont{Ewens}(1989)}]{ewens1989interpretation}
\bibinfo{author}{\bibfnamefont{W.~J.} \bibnamefont{Ewens}},
  \bibinfo{journal}{Theoretical population biology}
  \textbf{\bibinfo{volume}{36}}, \bibinfo{pages}{167} (\bibinfo{year}{1989}).

\bibitem[{\citenamefont{Crowley et~al.}(2005)\citenamefont{Crowley, Davis,
  Ensminger, Fuselier, Kasi~Jackson, and
  Nicholas~McLetchie}}]{crowley2005general}
\bibinfo{author}{\bibfnamefont{P.~H.} \bibnamefont{Crowley}},
  \bibinfo{author}{\bibfnamefont{H.~M.} \bibnamefont{Davis}},
  \bibinfo{author}{\bibfnamefont{A.~L.} \bibnamefont{Ensminger}},
  \bibinfo{author}{\bibfnamefont{L.~C.} \bibnamefont{Fuselier}},
  \bibinfo{author}{\bibfnamefont{J.}~\bibnamefont{Kasi~Jackson}},
  \bibnamefont{and}
  \bibinfo{author}{\bibfnamefont{D.}~\bibnamefont{Nicholas~McLetchie}},
  \bibinfo{journal}{Ecology Letters} \textbf{\bibinfo{volume}{8}},
  \bibinfo{pages}{176} (\bibinfo{year}{2005}).

\bibitem[{\citenamefont{Lloyd and Allen}(2015)}]{lloyd2015competition}
\bibinfo{author}{\bibfnamefont{D.~P.} \bibnamefont{Lloyd}} \bibnamefont{and}
  \bibinfo{author}{\bibfnamefont{R.~J.} \bibnamefont{Allen}},
  \bibinfo{journal}{Journal of The Royal Society Interface}
  \textbf{\bibinfo{volume}{12}}, \bibinfo{pages}{20150608}
  (\bibinfo{year}{2015}).

\bibitem[{\citenamefont{Haeno et~al.}(2013)\citenamefont{Haeno, Maruvka, Iwasa,
  and Michor}}]{haeno2013stochastic}
\bibinfo{author}{\bibfnamefont{H.}~\bibnamefont{Haeno}},
  \bibinfo{author}{\bibfnamefont{Y.~E.} \bibnamefont{Maruvka}},
  \bibinfo{author}{\bibfnamefont{Y.}~\bibnamefont{Iwasa}}, \bibnamefont{and}
  \bibinfo{author}{\bibfnamefont{F.}~\bibnamefont{Michor}},
  \bibinfo{journal}{PloS one} \textbf{\bibinfo{volume}{8}},
  \bibinfo{pages}{e65724} (\bibinfo{year}{2013}).

\bibitem[{\citenamefont{Danino et~al.}(2018)\citenamefont{Danino, Kessler, and
  Shnerb}}]{danino2016stability}
\bibinfo{author}{\bibfnamefont{M.}~\bibnamefont{Danino}},
  \bibinfo{author}{\bibfnamefont{D.~A.} \bibnamefont{Kessler}},
  \bibnamefont{and} \bibinfo{author}{\bibfnamefont{N.~M.}
  \bibnamefont{Shnerb}}, \bibinfo{journal}{Theoretical Population Biology}
  \textbf{\bibinfo{volume}{119}}, \bibinfo{pages}{57} (\bibinfo{year}{2018}).

\bibitem[{\citenamefont{Danino et~al.}(2016)\citenamefont{Danino, Shnerb,
  Azaele, Kunin, and Kessler}}]{danino2016effect}
\bibinfo{author}{\bibfnamefont{M.}~\bibnamefont{Danino}},
  \bibinfo{author}{\bibfnamefont{N.~M.} \bibnamefont{Shnerb}},
  \bibinfo{author}{\bibfnamefont{S.}~\bibnamefont{Azaele}},
  \bibinfo{author}{\bibfnamefont{W.~E.} \bibnamefont{Kunin}}, \bibnamefont{and}
  \bibinfo{author}{\bibfnamefont{D.~A.} \bibnamefont{Kessler}},
  \bibinfo{journal}{Journal of theoretical biology}
  \textbf{\bibinfo{volume}{409}}, \bibinfo{pages}{155} (\bibinfo{year}{2016}).

\bibitem[{\citenamefont{Hidalgo et~al.}(2017)\citenamefont{Hidalgo, Suweis, and
  Maritan}}]{hidalgo2017species}
\bibinfo{author}{\bibfnamefont{J.}~\bibnamefont{Hidalgo}},
  \bibinfo{author}{\bibfnamefont{S.}~\bibnamefont{Suweis}}, \bibnamefont{and}
  \bibinfo{author}{\bibfnamefont{A.}~\bibnamefont{Maritan}},
  \bibinfo{journal}{Journal of theoretical biology}
  \textbf{\bibinfo{volume}{413}}, \bibinfo{pages}{1} (\bibinfo{year}{2017}).

\bibitem[{\citenamefont{Crow et~al.}(1970)\citenamefont{Crow, Kimura
  et~al.}}]{crow1970introduction}
\bibinfo{author}{\bibfnamefont{J.~F.} \bibnamefont{Crow}},
  \bibinfo{author}{\bibfnamefont{M.}~\bibnamefont{Kimura}},
  \bibnamefont{et~al.}, \bibinfo{journal}{An introduction to population
  genetics theory.}  (\bibinfo{year}{1970}).

\bibitem[{\citenamefont{Gerrish and Lenski}(1998)}]{gerrish1998fate}
\bibinfo{author}{\bibfnamefont{P.~J.} \bibnamefont{Gerrish}} \bibnamefont{and}
  \bibinfo{author}{\bibfnamefont{R.~E.} \bibnamefont{Lenski}},
  \bibinfo{journal}{Genetica} \textbf{\bibinfo{volume}{102}},
  \bibinfo{pages}{127} (\bibinfo{year}{1998}).

\bibitem[{\citenamefont{Danino and Shnerb}(2018)}]{danino2017fixation}
\bibinfo{author}{\bibfnamefont{M.}~\bibnamefont{Danino}} \bibnamefont{and}
  \bibinfo{author}{\bibfnamefont{N.~M.} \bibnamefont{Shnerb}},
  \bibinfo{journal}{arXiv preprint arXiv:1710.08807. Journal of Theoretical
  Biology (in press)}  (\bibinfo{year}{2018}).

\bibitem[{\citenamefont{Meyer and Shnerb}(2018)}]{meyer2018noise}
\bibinfo{author}{\bibfnamefont{I.}~\bibnamefont{Meyer}} \bibnamefont{and}
  \bibinfo{author}{\bibfnamefont{N.~M.} \bibnamefont{Shnerb}},
  \bibinfo{journal}{arXiv preprint arXiv:1801.05970}  (\bibinfo{year}{2018}).

\bibitem[{\citenamefont{Fisher}(1937)}]{fisher1937wave}
\bibinfo{author}{\bibfnamefont{R.~A.} \bibnamefont{Fisher}},
  \bibinfo{journal}{Annals of Human Genetics} \textbf{\bibinfo{volume}{7}},
  \bibinfo{pages}{355} (\bibinfo{year}{1937}).

\bibitem[{\citenamefont{Van~Saarloos}(2003)}]{van2003front}
\bibinfo{author}{\bibfnamefont{W.}~\bibnamefont{Van~Saarloos}},
  \bibinfo{journal}{Physics reports} \textbf{\bibinfo{volume}{386}},
  \bibinfo{pages}{29} (\bibinfo{year}{2003}).

\bibitem[{\citenamefont{Maruvka and Shnerb}(2006)}]{maruvka2006nonlocal}
\bibinfo{author}{\bibfnamefont{Y.~E.} \bibnamefont{Maruvka}} \bibnamefont{and}
  \bibinfo{author}{\bibfnamefont{N.~M.} \bibnamefont{Shnerb}},
  \bibinfo{journal}{Physical Review E} \textbf{\bibinfo{volume}{73}},
  \bibinfo{pages}{011903} (\bibinfo{year}{2006}).

\bibitem[{\citenamefont{Fuentes et~al.}(2003)\citenamefont{Fuentes, Kuperman,
  and Kenkre}}]{fuentes2003nonlocal}
\bibinfo{author}{\bibfnamefont{M.}~\bibnamefont{Fuentes}},
  \bibinfo{author}{\bibfnamefont{M.}~\bibnamefont{Kuperman}}, \bibnamefont{and}
  \bibinfo{author}{\bibfnamefont{V.}~\bibnamefont{Kenkre}},
  \bibinfo{journal}{Physical review letters} \textbf{\bibinfo{volume}{91}},
  \bibinfo{pages}{158104} (\bibinfo{year}{2003}).

\bibitem[{\citenamefont{Hekstra and Leibler}(2012)}]{hekstra2012contingency}
\bibinfo{author}{\bibfnamefont{D.~R.} \bibnamefont{Hekstra}} \bibnamefont{and}
  \bibinfo{author}{\bibfnamefont{S.}~\bibnamefont{Leibler}},
  \bibinfo{journal}{Cell} \textbf{\bibinfo{volume}{149}}, \bibinfo{pages}{1164}
  (\bibinfo{year}{2012}).

\bibitem[{\citenamefont{Chisholm et~al.}(2014)\citenamefont{Chisholm, Condit,
  Rahman, Baker, Bunyavejchewin, Chen, Chuyong, Dattaraja, Davies, Ewango
  et~al.}}]{chisholm2014temporal}
\bibinfo{author}{\bibfnamefont{R.~A.} \bibnamefont{Chisholm}},
  \bibinfo{author}{\bibfnamefont{R.}~\bibnamefont{Condit}},
  \bibinfo{author}{\bibfnamefont{K.~A.} \bibnamefont{Rahman}},
  \bibinfo{author}{\bibfnamefont{P.~J.} \bibnamefont{Baker}},
  \bibinfo{author}{\bibfnamefont{S.}~\bibnamefont{Bunyavejchewin}},
  \bibinfo{author}{\bibfnamefont{Y.-Y.} \bibnamefont{Chen}},
  \bibinfo{author}{\bibfnamefont{G.}~\bibnamefont{Chuyong}},
  \bibinfo{author}{\bibfnamefont{H.}~\bibnamefont{Dattaraja}},
  \bibinfo{author}{\bibfnamefont{S.}~\bibnamefont{Davies}},
  \bibinfo{author}{\bibfnamefont{C.~E.} \bibnamefont{Ewango}},
  \bibnamefont{et~al.}, \bibinfo{journal}{Ecology letters}
  \textbf{\bibinfo{volume}{17}}, \bibinfo{pages}{855} (\bibinfo{year}{2014}).

\bibitem[{\citenamefont{Kalyuzhny
  et~al.}(2014{\natexlab{a}})\citenamefont{Kalyuzhny, Seri, Chocron, Flather,
  Kadmon, and Shnerb}}]{kalyuzhny2014niche}
\bibinfo{author}{\bibfnamefont{M.}~\bibnamefont{Kalyuzhny}},
  \bibinfo{author}{\bibfnamefont{E.}~\bibnamefont{Seri}},
  \bibinfo{author}{\bibfnamefont{R.}~\bibnamefont{Chocron}},
  \bibinfo{author}{\bibfnamefont{C.~H.} \bibnamefont{Flather}},
  \bibinfo{author}{\bibfnamefont{R.}~\bibnamefont{Kadmon}}, \bibnamefont{and}
  \bibinfo{author}{\bibfnamefont{N.~M.} \bibnamefont{Shnerb}},
  \bibinfo{journal}{The American Naturalist} \textbf{\bibinfo{volume}{184}},
  \bibinfo{pages}{439} (\bibinfo{year}{2014}{\natexlab{a}}).

\bibitem[{\citenamefont{Kalyuzhny
  et~al.}(2014{\natexlab{b}})\citenamefont{Kalyuzhny, Schreiber, Chocron,
  Flather, Kadmon, Kessler, and Shnerb}}]{kalyuzhny2014temporal}
\bibinfo{author}{\bibfnamefont{M.}~\bibnamefont{Kalyuzhny}},
  \bibinfo{author}{\bibfnamefont{Y.}~\bibnamefont{Schreiber}},
  \bibinfo{author}{\bibfnamefont{R.}~\bibnamefont{Chocron}},
  \bibinfo{author}{\bibfnamefont{C.~H.} \bibnamefont{Flather}},
  \bibinfo{author}{\bibfnamefont{R.}~\bibnamefont{Kadmon}},
  \bibinfo{author}{\bibfnamefont{D.~A.} \bibnamefont{Kessler}},
  \bibnamefont{and} \bibinfo{author}{\bibfnamefont{N.~M.}
  \bibnamefont{Shnerb}}, \bibinfo{journal}{Ecology}
  \textbf{\bibinfo{volume}{95}}, \bibinfo{pages}{1701}
  (\bibinfo{year}{2014}{\natexlab{b}}).

\bibitem[{\citenamefont{Fung et~al.}(2016)\citenamefont{Fung, O'Dwyer, Rahman,
  Fletcher, and Chisholm}}]{fung2016reproducing}
\bibinfo{author}{\bibfnamefont{T.}~\bibnamefont{Fung}},
  \bibinfo{author}{\bibfnamefont{J.~P.} \bibnamefont{O'Dwyer}},
  \bibinfo{author}{\bibfnamefont{K.~A.} \bibnamefont{Rahman}},
  \bibinfo{author}{\bibfnamefont{C.~D.} \bibnamefont{Fletcher}},
  \bibnamefont{and} \bibinfo{author}{\bibfnamefont{R.~A.}
  \bibnamefont{Chisholm}}, \bibinfo{journal}{Ecology}
  \textbf{\bibinfo{volume}{97}}, \bibinfo{pages}{1207} (\bibinfo{year}{2016}).

\bibitem[{\citenamefont{Wienand et~al.}(2017)\citenamefont{Wienand, Frey, and
  Mobilia}}]{wienand2017evolution}
\bibinfo{author}{\bibfnamefont{K.}~\bibnamefont{Wienand}},
  \bibinfo{author}{\bibfnamefont{E.}~\bibnamefont{Frey}}, \bibnamefont{and}
  \bibinfo{author}{\bibfnamefont{M.}~\bibnamefont{Mobilia}},
  \bibinfo{journal}{Physical review letters} \textbf{\bibinfo{volume}{119}},
  \bibinfo{pages}{158301} (\bibinfo{year}{2017}).

\bibitem[{\citenamefont{Shtilerman et~al.}(2015)\citenamefont{Shtilerman,
  Kessler, and Shnerb}}]{shtilerman2015emergence}
\bibinfo{author}{\bibfnamefont{E.}~\bibnamefont{Shtilerman}},
  \bibinfo{author}{\bibfnamefont{D.~A.} \bibnamefont{Kessler}},
  \bibnamefont{and} \bibinfo{author}{\bibfnamefont{N.~M.}
  \bibnamefont{Shnerb}}, \bibinfo{journal}{Journal of theoretical biology}
  \textbf{\bibinfo{volume}{383}}, \bibinfo{pages}{138} (\bibinfo{year}{2015}).

\end{thebibliography}

\end{document}